\documentclass{egpubl}
\usepackage{DH2025}%
 
\WsPaper           

\usepackage[T1]{fontenc}
\usepackage{dfadobe}  

\usepackage{cite}  
\BibtexOrBiblatex
\electronicVersion
\PrintedOrElectronic
\ifpdf \usepackage[pdftex]{graphicx} \pdfcompresslevel=9
\else \usepackage[dvips]{graphicx} \fi

\usepackage{egweblnk} 

\title[ARise: an Augmented Reality Mobile Application to Improve Cultural Heritage Resilience]%
      {ARise: an Augmented Reality Mobile Application to Improve Cultural Heritage Resilience}

\author[Urbanelli et al.]
{\parbox{\textwidth}{\centering Angelica Urbanelli$^{1}$\orcid{0000-0003-3403-4367}, 
         Marina Nadalin$^{1}$\orcid{0009-0006-9250-8005}, 
         Mario Chiesa$^{1}$\orcid{0000-0001-8719-380X}, 
         Rojin Bayat$^{1}$\orcid{0009-0004-8558-3854}, 
         Massimo Migliorini$^{1}$\orcid{0000-0001-9966-5026}, 
         Claudio Rossi$^{1}$\orcid{0000-0001-5038-3597} 
        }
        \\
{\parbox{\textwidth}{\centering $^1$ LINKS Foundation, Torino (TO), Italy\\
       }
}
}

\begin{document}


\maketitle
\begin{abstract}
   The preservation of cultural heritage faces increasing threats from climate change effects and environmental hazards, demanding innovative solutions that can promote awareness and resilience. This paper presents ARise, an Augmented Reality mobile application designed to enhance public engagement with cultural sites while raising awareness about the local impacts of climate change. Based on a user-centered co-creative methodology involving stakeholders from five European regions, ARise integrates multiple data sources\textemdash a Crowdsourcing Chatbot, a Social Media Data Analysis tool, and an AI-based Artwork Generation module\textemdash to deliver immersive and emotionally engaging experiences. Although formal user testing is forthcoming, this prototype demonstrates the potential of AR to support education, cultural sustainability, and climate adaptation. \\


\end{abstract}  

\section{Introduction}

Cultural Heritage (CH) is a fundamental pillar of our communities and economies, shaping our shared identity and fostering social cohesion. However, this heritage faces increasing threats due to climate change and natural hazards, putting its preservation at risk. 
In this context, emerging technologies\textemdash particularly Augmented Reality (AR)\textemdash may offer new opportunities to effectively promote cultural sites in an educational and entertaining way, and create stronger interactions with residents and tourists, activating virtuous and effective behaviors. In this way, users can interact with Points of Interest (PoIs) in a meaningful and immersive experience that connect them with CH, improving awareness, emotional connections, and supporting heritage resilience in multiple innovative ways \cite{INNOCENTE2023268}.

In this paper we present ARise, an AR-based application designed to support CH preservation and resilience and raise awareness about climate change's effects in cultural sites. The application enriches the user experience by integrating data collected and processed through other tools, 
namely a Crowdsourcing Chatbot, a Social Media Data Analysis Tool (SMDA), and an Artificial Intelligence (AI) Artworks Generation module, each contributing to the creation of a rich, interactive experience that deepens users' engagement. While the app has been fully developed and tested internally, it has not yet been evaluated with end-users; future work will focus on a thorough evaluation and refinement based on user tests.

\section{Related works}

The preservation of CH has long been a central concern across multiple disciplines, including architecture, archaeology, conservation, and digital technology. In recent years, this focus has expanded to include the ways in which emerging technologies can support cultural resilience in the face of growing environmental and social pressures. Among these technologies, AR stands out as a particularly promising tool as it offers innovative opportunities for documenting, preserving, and promoting heritage sites through interactive and immersive digital content \cite{Ibis2024Sus}. Additionally, 
the integration of AR into the CH domain aligns with a broader shift towards more dynamic and participatory models of heritage engagement. Bekele et al. \cite{Bekele2018} provide a comprehensive review of AR applications in this field, identifying its capacity to support the digital preservation of tangible and intangible heritage while enhancing user experience through enriched storytelling and spatial contextualization. Their work underscores how AR can create emotionally resonant and cognitively engaging encounters with heritage that go beyond passive observation.
At the same time, the role of AR has been further explored in participatory frameworks that aim to empower local communities towards CH. A study from Kwiecinski et al. \cite{Kwiecinski2017DesignAR} highlights the importance of involving citizens in the co-design of AR tools for heritage, emphasizing how digital experiences can encourage and motivate people to actively support preservation of cultural sites. These contributions position AR not just as a display technology, but as a catalyst for civic engagement and adaptive heritage governance. 
Another emerging technology recently applied in the CH domain is AI, which is being adopted for a wide range of purposes\textemdash from the digital reconstruction of damaged artifacts to the development of intelligent interfaces and educational tools \cite{MISHRA2024536}. Among the various branches of AI, generative AI has recently gained particular celebrity, with tools such as ChatGPT, Gemini, and DALL\textperiodcentered E capturing global attention and transforming how cultural content can be created, interpreted, and perceived.
Recent research has begun to explore how generative AI can contribute to the valorization of intangible CH. Zhang et al. \cite{zhang2023can} examined the use of AI to generate cultural and creative products inspired by the Chinese tradition of New Year Prints, revealing a strong positive correlation between users' attraction, interest, and participation in AI-generated artworks and their perceived value. The study highlights how AI-generated cultural content can engage the public, particularly younger audiences, by offering novel artistic expressions, reinforcing their connection to cultural identity.

\section{System Design and Methodology}

The design and development of the ARise app is part of a wider project consisting in an ecosystem of digital tools co-created to protect local CH and landscape from the effects of climate change. In the context of the project, five stakeholders from five different European regions (Italy, Spain, Greece, Germany and Croatia) have been involved to take part in the co-creation process of these tools, including the ARise app. As representatives from three UNESCO sites, one FAO GIAHS site, and one National Park, they played a crucial role in highlighting the diversities of cultural heritage, along with risks and vulnerabilities. 
This collaborative and interdisciplinary approach enabled the design and development of tools that are not limited to a single case study but are instead adaptable across different local needs and scenarios. 

Following these objectives, we elaborated and implemented a user-centered design methodology inspired by the principles of Design Thinking \cite{DesignThinking}, identifying user focus and problem framing though several workshops with the five stakeholders involved. 
The identified methodology has been organized into five main phases: (i) collecting data and knowledge; (ii) understanding risks and impacts; (iii) developing strategies and solutions; (iv) disseminating knowledge to targeted stakeholders; and (v) informing people to promote awareness and engagement. 
As part of the initial phase of data collection, we gathered needs and requirements and grouped them in four categories: (1) \textit{UX/UI requirements}, regarding the overall user experience and elements of the user interface; (2) \textit{Functional}, like specific functionalities and features; (3) \textit{Content}, e.g. text, data and multimedia; (4) \textit{Deployment}, technical aspects such as integrations with existing platforms or compatibility with standards.\par
\begin{figure}[htb]
  \centering  \includegraphics[width=\linewidth]{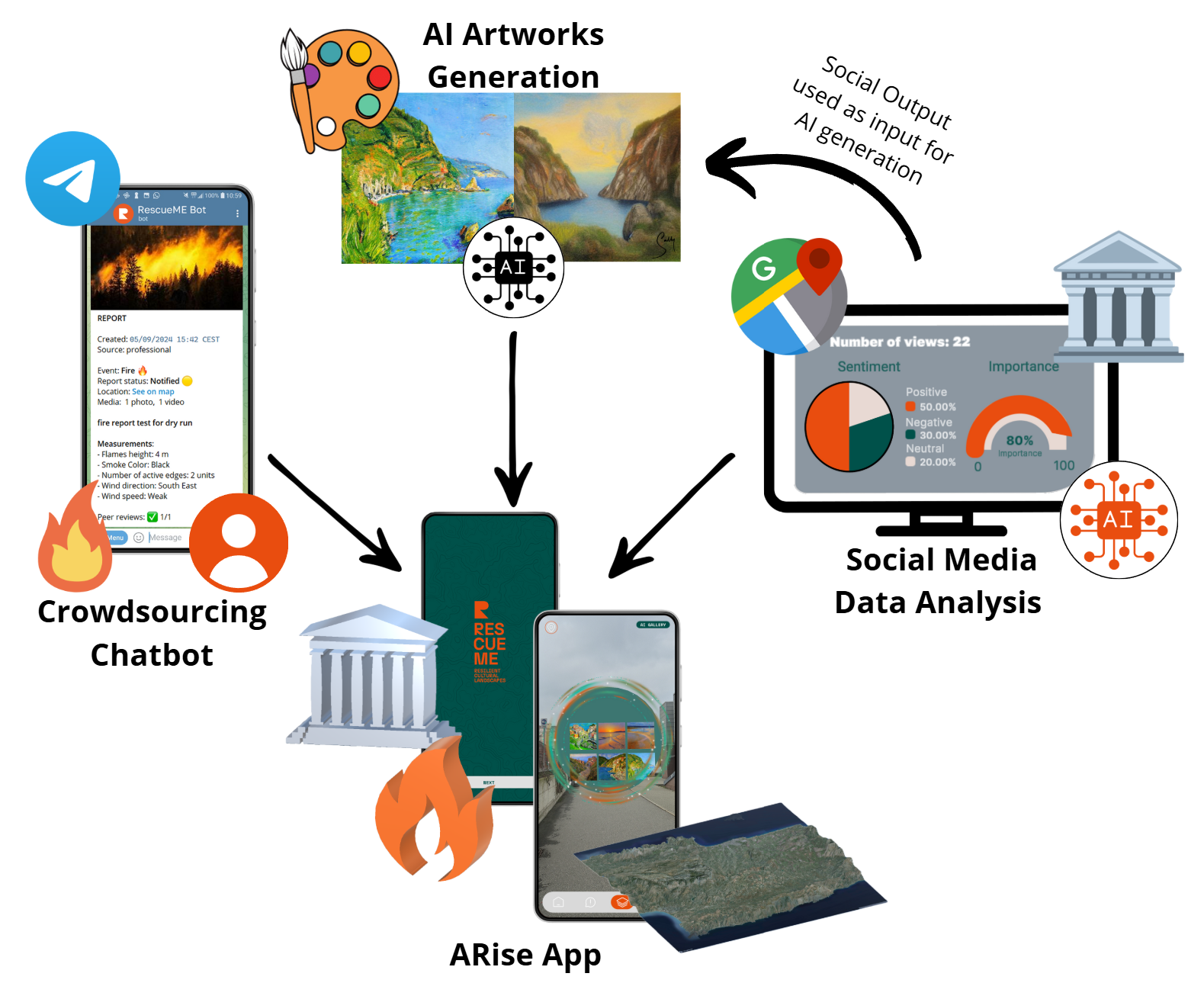}
  \caption{Overall system's schema representing connections between the tools and the ARise App}
  \label{fig:ToolsSchema}
\end{figure}

  
The conceptual foundation of the app is based on the idea that technology can support climate resilience, especially when it comes to protecting cultural and natural heritage. In a context where communities are increasingly exposed to the impacts of climate change, the app encourages a shift in perspective: adaptation is not only a matter of infrastructure or policy, but also a personal and collective process that involves awareness, emotional engagement, and the capacity to act.
The collected design insights and user requirements revealed not only informational gaps but also a general emotional distance between users and the abstract data on environmental risks. The insights gathered were translated into a design concept that prioritized emotional resonance, narrative interaction, and the meaningful use of emerging technologies. This strategy is encapsulated in a user flow within the app itself that moves from seeing and understanding, to feeling, reflecting, and ultimately acting.
The resulting application leverages AR to present digital contents ensuring access to meaningful experiences regardless of users' physical location. These contents are generated and processed by the ecosystem of tools, and finally retrieved and displayed through the AR interface (Figure \ref{fig:ToolsSchema}). Although all the tools presented have been developed within the context of the project, the AR application stands out as the core contribution of this work, integrating and visualizing data from the other tools in a coherent and engaging way. 

While formal user evaluation is planned for future stages, the outcome of this work is a fully functional prototype that meets the technical requirements collected at the beginning of the design phase. A video of the application navigation flow has been uploaded into Zenodo platform and it is available at \href{https://zenodo.org/records/15853870}{ARise}. 

\subsection{AR App}
\label{AR App}

The ARise application is structured around a clear and modular navigation flow that begins with user authentication and leads to a personalized homepage, from which all functionalities are accessible. The application integrates multiple data sources and tools into a cohesive AR interface and offer a gamified system which tracks users interaction and progress, rewarding engagement and encouraging learning. The application is built using Unity and AR Foundation, ensuring platform compatibility, real-time spatial tracking, and a seamless user experience across devices. 
To define how data should be visualized in AR, a preliminary design phase analysed the nature, structure, and relevance of each dataset. This process informed the selection of trigger mechanisms (e.g., GPS or plane detection) and visualization formats (e.g., 3D models, charts, or galleries), ensuring that content would be spatially meaningful, accessible, and engaging. The main interface is organized into two core AR modes\textemdash OnSite and OffSite\textemdash allowing users to explore climate and cultural heritage data either through direct geolocated experiences or remotely, via spatially anchored visualizations.

\begin{figure}[htb]
  \centering  \includegraphics[width=0.75\linewidth]{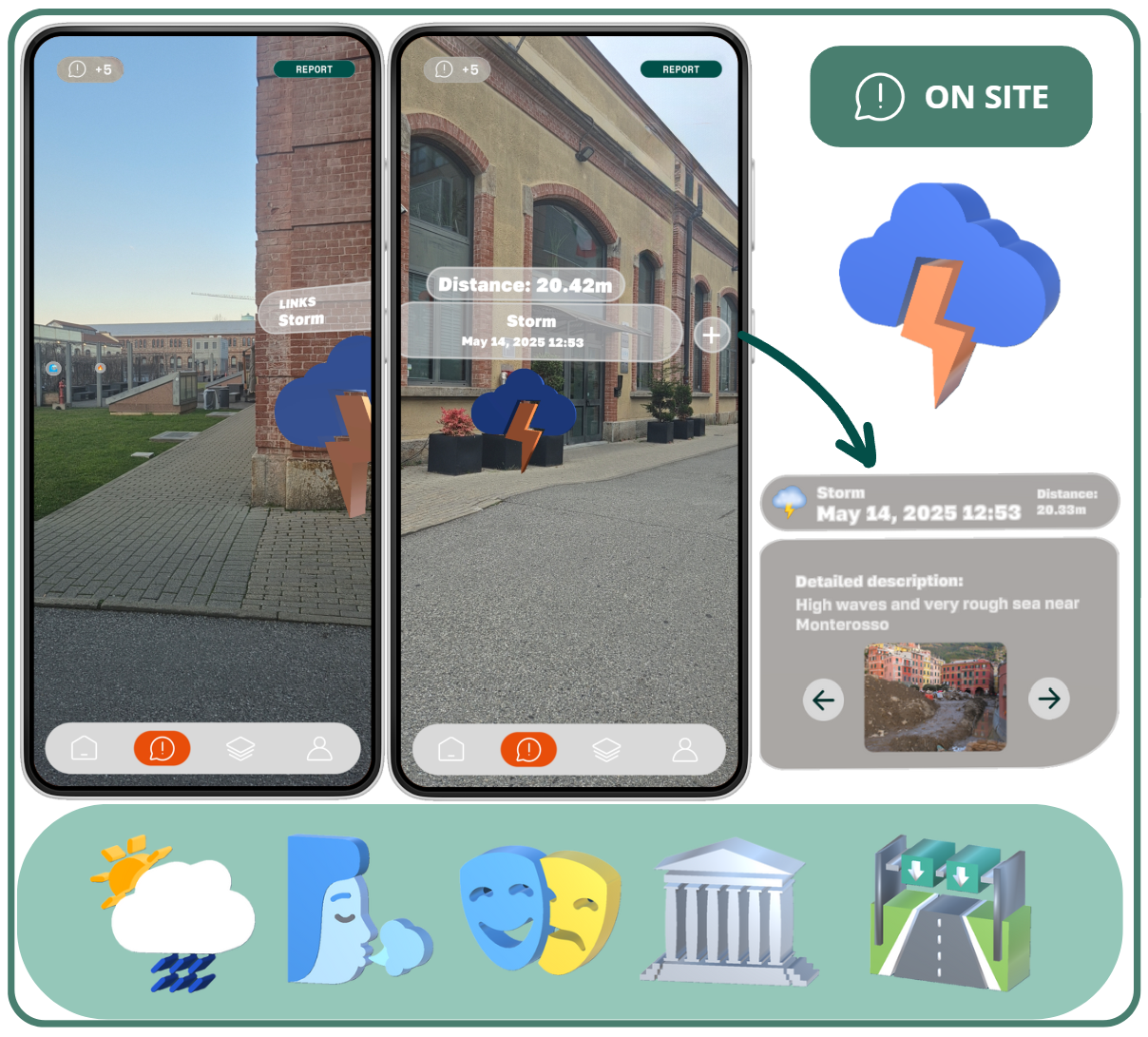}
  \caption{ARise OnSite Section: an example of how Chatbot reports are visualized using the corresponding 3D hazard model and a UI with detailed informations}
  \label{fig:onsite System}
\end{figure}

In the \textit{onsite mode}, the application uses GPS and plane detection to provide a location-based AR experience. When physically present at a site, users can access geolocated reports submitted 
in the Crowdsourcing Chatbot and the results of the SMDA in specific PoIs. These elements are visualized as 3D models placed in the real-world environment according to their corresponding geographic coordinates. Some examples of 3D models can be seen in the \textit{onsite} block of Figure \ref{fig:onsite System}.
The Chatbot reports are represented as 3D models corresponding to the type of hazard they describe (e.g., fire, flood, storm). When the model is selected, a popup card appears providing detailed information, such as a brief description, the distance from the user, and, when available, an image of the reported event.
Similarly, the PoIs' analysis results are visualized by geolocating the identified points with corresponding 3D model categories. When the user interacts with these models, an overlay appears displaying the SMDA results, namely the number of reviews, the sentiment analysis results, and the importance score of the site, described in more detail in Section \ref{Social Media Data Analysis Tool}.

The \textit{offsite mode} enables access to digital heritage content and environmental information even when users are not physically present at the target location. It includes three main functionalities: (1) \textit{3D data visualization} of social and territorial datasets via interactive graphs displayed on 3D maps; (2) \textit{Augmented reality galleries}, immersive virtual spaces presenting AI-generated artworks associated with CH sites; (3) \textit{Climate impact visualizations}, where users can explore 3D maps of different locations showing variations in environmental indicators. All three types of visualizations are activated using plane recognition: once the device detects a suitable surface, the corresponding content 
 is anchored in the physical space and becomes visible to the user. 
The 3D maps are generated from satellite imagery of the areas of interest and reconstructed using height maps, ensuring a realistic 
representation of the terrain. In these maps users can interact with some indicators, namely temperature, water levels and vegetation coverage, and see the potential effects of their variation. 
User interaction is thus designed to be engaging and exploratory, encouraging curiosity and deeper understanding of climate change effects. 
\begin{figure}[htb]
  \centering  \includegraphics[width=0.75\linewidth]{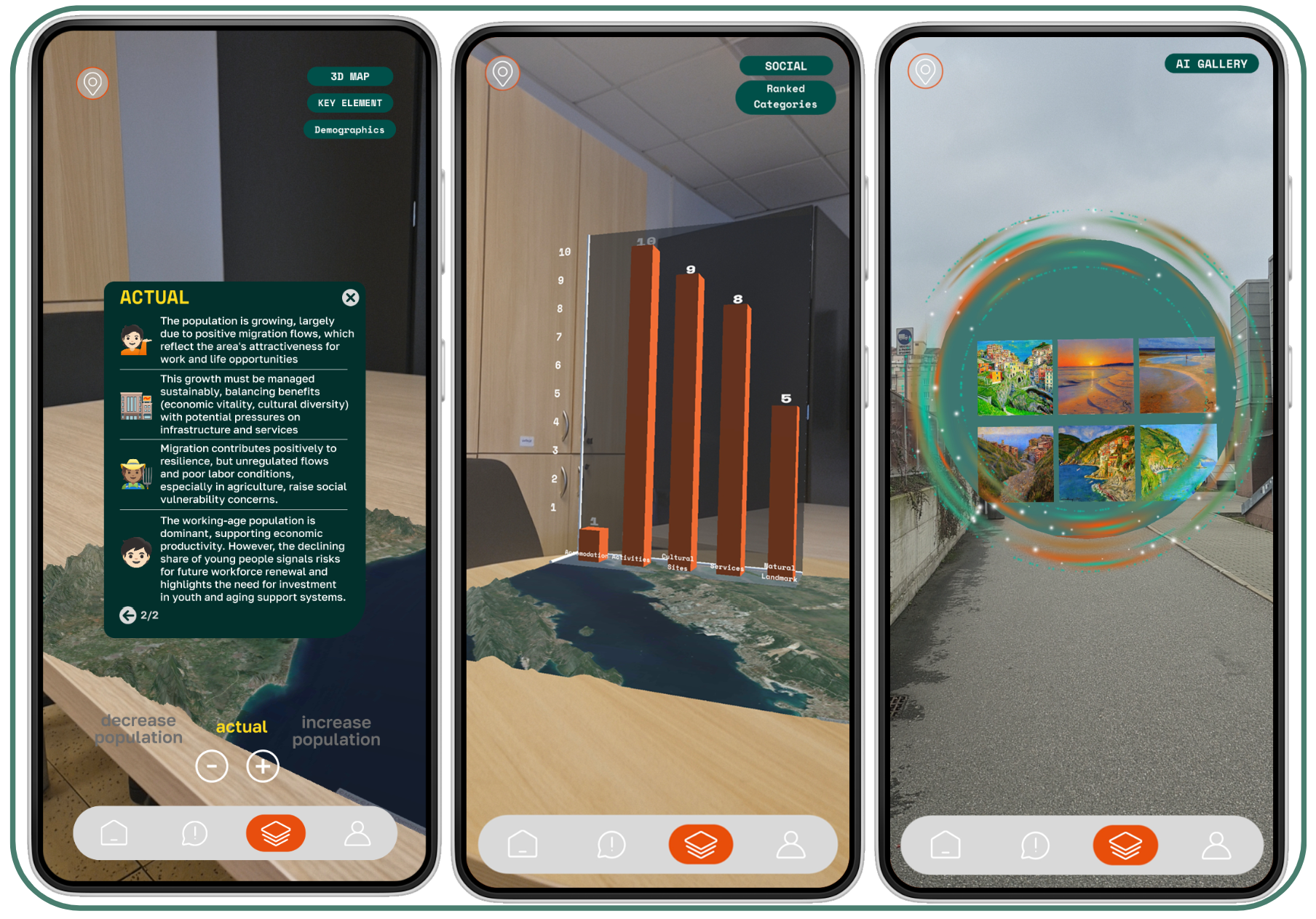}
  \caption{ARise Offsite Section and its three functionalities: Climate impact visualization, 3D data visualization of social datasets both displayed on 3D maps and immersive AR portal, gateway to the AI-artworks gallery}
  \label{fig:offsite System}
\end{figure}
\vspace{-5mm}
\subsection{Social Media Data Analysis Tool}
\label{Social Media Data Analysis Tool}
As an integrated data source for the ARise App, the SMDA provides innovative means to understand and monitor public perception of CH and landscapes. The aim of this tool is to implement an automated data processing flow, to identify trends and sentiment patterns associated with specific cultural sites, enabling users to gain valuable insights into public opinions from other visitors and citizens, and supporting professionals for evidence-based decision-making and strategy development.
The analysis of this module is performed on data collected regarding specific relevant PoIs, identified in collaboration with the stakeholders of the five different case studies.
The sources used by this tool to extract and elaborate data are: (1) user reviews from Google Maps, analysed to perform sentiment analysis and compute an importance score\textemdash the former assigns a numerical score to each review, while the latter combines sentiment scores with the number of reviews, to reflect the relevance of a cultural site; (2) Google Maps images, for automated extraction of descriptions, useful for further informative insights. 
The output generated by SMDA is used for (a) the generation of images through Generative AI (Section \ref{AI Artworks Generation}) and (b) data visualization in the AR App, as previously described in Section \ref{AR App}. 

\subsection{Crowdsourcing Chatbot}
\label{Crowdsourcing Chatbot}
The ARise App displays geolocated reports submitted by citizens through a Crowdsourcing Chatbot. The methodology adopted for the collection of requirements, as well as the design, development, and user testing of this tool, is extensively described in a previous work \cite{urbanelli2024ermes}. This tool is a Telegram-based Chatbot that allows users to share with local authorities real-time updates about what is happening nearby, including potential hazards and at-risk or damaged cultural sites. To submit a report, users simply need to follow the Chatbot's instructions, providing the required information\textemdash including GPS location, type of hazard, a textual description, and various kinds of multimedia content such as photos, videos, or voice messages\textemdash which can be recorded and shared directly through the chat interface. In addition to the efforts carried out in previous projects, this work introduced a co-creation process aimed at aligning the report creation workflow with end users' needs. To this end, dedicated workshops and co-design sessions has been organized with the five involved stakeholders to identify which additional information would be most valuable to include in the reports, based on their real-world use cases. The outcome of this process is a detailed list of measurement types, impact indicators, and risk assessment elements that stakeholders explicitly requested and finally incorporated into the report creation process.  
\subsection{AI Artworks Generation}
\label{AI Artworks Generation}

Given the recent growing public interest in generative AI, we decided to dedicate one section of the app to create an immersive experience where users could see and interact with AI-generated artworks as if they are inside a virtual museum. The goal is to create an experience that can engage tourists and local citizens, while also promoting the local CH, such as landscapes and cultural sites. To achieve this, we developed a tool that leverages a generative AI model to transform photographs of selected PoIs into artistic images that reflects the sentiment expressed by end users. In this way, each generated painting represents not only the visual characteristics of the site, but also the emotional tone expressed by visitors through their reviews, resulting in a unique and dynamic artistic interpretation of the territory.
The AI model selected for this tool is Stable Diffusion v1.5, a latent diffusion model based on the architecture introduced by Rombach et al. \cite{Rombach_2022_CVPR}, which enables high-resolution image synthesis through a compressed latent space. Specifically, the model employed is pretrained on the LAION-5B dataset and released by CompVis and RunwayML \cite{stablediffusion15}.
The generative artwork tool is designed to be strictly integrated with the SMDA. As mentioned above, the SMDA performs the retrieval and analysis of Google Maps reviews and images. Once the analysis is performed, the AI Artworks Generation Tool retrieves periodically the top five most reviewed PoIs for each use case, along with the computed sentiment scores, and their photographs. For these selected locations, the corresponding sentiment scores are used as a prompt condition to generate new unique AI artworks starting from real sites' photographs. 

\section{Conclusions and Future Work}

In conclusion, this work aims to present the design and development of an innovative mobile application that leverages different emerging technologies in order to raise awareness regarding climate changes effects on CH, and promote resilience in local communities. 
The application is based on AR technology to enrich the physical world with interactive digital elements. 
By presenting environmental and social data in an engaging and immersive way, the application becomes an interactive information tool that enables users not only to understand real-world problems but also to act upon them, thereby enhancing their individual and collective resilience in relation to heritage and landscape preservation. 
Although the application has been fully developed and tested internally, future work will focus on conducting user tests to evaluate the system's usability and effectiveness in meeting end-user needs. These evaluations will provide valuable feedback for refining specific features and functionalities, enabling further fine-tuning of the application based on user tests outcome.


\section*{Acknowledgements}
This work is partially funded by the European Commission through the Horizon Europe Programme under the following project: RescueME - equitable RESilience solutions to strengthen the link between CULtural landscapes and coMmunitiEs, Call HORIZON-CL2-2022-HERITAGE-01-08, Grant Agreement n. 101094978.
\bibliographystyle{eg-alpha-doi} 
\bibliography{egbibv2}       




\end{document}